\documentclass[prb,twocolumn]{revtex4}%

\usepackage{graphicx}
\usepackage{amsmath}
\usepackage{ulem}
\usepackage{color}

\newcommand{\be}{\begin{equation}}
\newcommand{\ee}{\end{equation}}
\newcommand{\bea}{\begin{eqnarray*}}
\newcommand{\eea}{\end{eqnarray*}}
\newcommand{\bean}{\begin{eqnarray}}
\newcommand{\eean}{\end{eqnarray}}
\def\lan{\langle}
\def\ran{\rangle}

\begin{document}

\draft
\title
{\bf A heat engine made of quantum dot molecules with high figure of
merits}

\author{Chih-Chieh Chen,$^{1}$ David M T Kuo$^{2}$ and Yia-Chung Chang$^{1,3,4}$}
\address{$^1$Department of Physics, University of Illinois at Urbana-Champaign, Urbana, Illinois 6180, USA}

\address{$^2$Department of Electrical Engineering and Department of Physics, National Central
University, Chungli, 320 Taiwan}

\address{$^{3}$Research Center for Applied Sciences, Academic Sinica,
Taipei, 11529 Taiwan} \affiliation{$^4$ Department of Physics,
National Cheng Kung University, Tainan, 701 Taiwan}

\date{\today}

\begin{abstract}
The transport of electrons through serially coupled quantum dot
molecules (SCQDM) is investigated theoretically for application as
an energy harvesting engine (EHE), which converts thermal heat to
electrical power. We demonstrate that the charge current driven by a
temperature bias shows bipolar oscillatory behavior with respect to
gate voltage due to {the unbalance between electrons and
holes}, which is different from the charge current driven by an
applied bias. In addition, we reveal a Lenz's law between the charge
current and the thermal induced voltage. The efficiency of EHE is
higher for SCQDM in the orbital depletion situation rather than the
orbital filling situation, owing to the many-body effect. The EHE
efficiency is enhanced with increasing temperature bias, but
suppressed as the electron hopping strength reduces. The fluctuation
of QD energy levels at different sites also leads to a reduction of
EHE efficiency. Finally, we demonstrate direction-dependent charge
currents driven by the temperature bias for application as a novel
charge diode.
\end{abstract}

\maketitle
\section{Introduction}
Energy harvesting of heat dissipated from electronic circuits and
other heat sources is one of the most important energy issues.[1]
The realization of such type of energy harvesting typically relies
on the search of thermoelectric (TE) materials with high figure of
merits ($ZT$).[2] Impressive ZT values for quantum-dot superlattices
(QDSL) systems have been demonstrated experimentally.[3] The
enhancement of ZT mainly arises from the reduction of phonon thermal
conductivity in QDSL, which is due to the increased rate of phonon
scattering from the interface of quantum dots (QDs).[1,2] If the ZT
value can reach 3, the solid state cooler will have the potential to
replace conventional compressor-based air conditioners owing to its
long life time, low noise and low air pollution. Besides the search
of TE devices with large ZT value, the optimizing of nonlinear
thermoelectric behavior under high temperature bias is crucial for
the design of the next-generation energy harvesting engine
(EHE).[1,2]

Recently, a grate deal of efforts was devoted to the studies of the
nonlinear response of thermoelectric devices under high temperature
bias. The nonlinear phonon flow of nanostructures with respect to
large temperature bias were investigated experimentally[4] and
theoretically.[5-8] The phonon thermal rectification behavior of
silicon nanowire (which has a very low efficiency) was reported
experimentally.[4] More recently, the highly efficient electron
thermal diode was reported in a superconductor junction system.[9]
However, such a thermal rectification behavior only exists at very
low temperatures. Unlike heat rectifiers which are used to control
the direction of heat flow [4-9], the design of an {EHE}
driven by a large temperature bias needs to optimize the efficiency
in the energy transfer from the waste heat[1,2]. To design a
nanoscale EHE, which can be integrated with semiconductor electronic
circuits, it is important not only to collect the waste heat but
also to improve the performance of electronic circuits.

So far, experimental studies of EHE made of semiconductor QD
molecules (QDMs) have not been reported, mainly due to technical
difficulties and the lack of theoretical designs. Therefore, it is
desirable to have theoretical studies which can provide useful
guidelines for the advancement of nanoscale TE technology. Most
theoretical studies of TE properties are limited in the linear
response regime.[10-13] The many-body effect of QDMs also presents a
big challenge to the development of theoretical studies for TE
properties. In this article, we study the nonlinear behavior of EHE
made of serially coupled triple QDs (SCTQDs) based on a previously
developed numerically method,[12,13] which can suitably address the
many body effect in the Coulomb blockade regime. In addition, we
investigate an engine with direction-dependent electrical output
driven by a temperature-bias for application as a novel nonlinear TE
devices.

\section{Formalism}
The inset of Fig. 1(a) shows the QD molecule (QDM) connected to two
metallic electrodes, one is in thermal contact with the heat source
at temperature $T_H$ (hot side) and the other with the heat sink
kept at temperature $T_C$ (cold side).  The heat flows from the hot
side through the QDM into the cold side. To reveal the charge and
heat currents driven by the temperature bias, we consider the
following Hamiltonian $H=H_0+H_{QD}$ for a SCTQDs:

\begin{eqnarray}
H_0& = &\sum_{k,\sigma} \epsilon_k
a^{\dagger}_{k,\sigma}a_{k,\sigma}+ \sum_{k,\sigma} \epsilon_k
b^{\dagger}_{k,\sigma}b_{k,\sigma}\\ \nonumber &+&\sum_{k,\sigma}
V_{k,L}d^{\dagger}_{L,\sigma}a_{k,\sigma}
+\sum_{k,\sigma}V_{k,R}d^{\dagger}_{R,\sigma}b_{k,\sigma}+c.c
\end{eqnarray}
where the first two terms describe the free electron gas of left and
right electrodes (hot and cold sides). $a^{\dagger}_{k,\sigma}$
($b^{\dagger}_{k,\sigma}$) creates  an electron of momentum $k$ and
spin $\sigma$ with energy $\epsilon_k$ in the left (right)
electrode. $V_{k,\ell}$ ($\ell=L,R$) describes the coupling between
the electrodes and the left (right) QD. $d^{\dagger}_{\ell,\sigma}$
($d_{\ell,\sigma}$) creates (destroys) an electron in the $\ell$-th
dot.

\begin{small}
\begin{eqnarray}
H_{QD}&=& \sum_{\ell,\sigma} E_{\ell} n_{\ell,\sigma}+
\sum_{\ell} U_{\ell} n_{\ell,\sigma} n_{\ell,\bar\sigma}\\
\nonumber &+&\frac{1}{2}\sum_{\ell,j,\sigma,\sigma'}
U_{\ell,j}n_{\ell,\sigma}n_{j,\sigma'}
+\sum_{\ell,j,\sigma}t_{\ell,j} d^{\dagger}_{\ell,\sigma}
d_{j,\sigma},
\end{eqnarray}
\end{small}
where { $E_{\ell}$} is the spin-independent QD energy level, and
$n_{\ell,\sigma}=d^{\dagger}_{\ell,\sigma}d_{\ell,\sigma}$.
Notations $U_{\ell}$ and $U_{\ell,j}$ describe the intradot and
interdot Coulomb interactions, respectively. $t_{\ell,j}$ describes
the electron interdot hopping. Noting that the interdot Coulomb
interactions as well as intradot Coulomb interactions play a
significant role on the charge transport in semiconductor QD arrays
or molecular chains.[10-13] Because we are interested in the case
that the thermal energy is much smaller than intradot Coulomb
interactions, we consider QDs with only one energy level per dot.

Using the Keldysh-Green's function technique,[14,15] the charge and
heat currents from reservoir $\alpha$ to the QDM junction are
calculated according to the Meir-Wingreen formula
\begin{eqnarray}
J_{\alpha}&=&\frac{ie}{h}\sum_{j\sigma}\int {d\epsilon}
\Gamma^\alpha_{j}(\epsilon) [ G^{<}_{j\sigma} (\epsilon)+ f_\alpha
(\epsilon)( G^{r}_{j\sigma}(\epsilon) \nonumber \\ &-&
G^{a}_{j\sigma}(\epsilon) ) ]\\ Q_{\alpha}
&=&\frac{i}{h}\sum_{j\sigma}\int {d\epsilon} (\epsilon-\mu_{\alpha})
\Gamma^\alpha_{j}(\epsilon) [ G^{<}_{j\sigma} (\epsilon) f_\alpha
(\epsilon) \nonumber \\ & &( G^{r}_{j\sigma}(\epsilon) -
G^{a}_{j\sigma}(\epsilon) ) ],
\end{eqnarray}
Notation $\Gamma^\alpha_{\ell}=\sum_k
|V_{k,L(R),\ell}|^2\delta(\epsilon-\epsilon_k)$ is the tunneling
rate between the left (right) reservoir and the left (right) QD of
QDM.
$f_{\alpha}(\epsilon)=1/\{\exp[(\epsilon-\mu_{\alpha})/k_BT_{\alpha}]+1\}$
denotes the Fermi distribution function for the $\alpha$-th
electrode, where $\mu_\alpha$  and $T_{\alpha}$ are the chemical
potential and the temperature of the $\alpha$ electrode.
$\mu_L-\mu_R=e\Delta V$ and $T_L-T_R=\Delta T$. $e$, $h$, and $k_B$
denote the electron charge, the Planck's constant, and the Boltzmann
constant, respectively. $G^{<}_{j\sigma} (\epsilon)$,
$G^{r}_{j\sigma}(\epsilon)$, and $G^{a}_{j\sigma}(\epsilon)$ are the
frequency domain representations of the one-particle lesser,
retarded, and advanced Green's functions
$G^{<}_{j\sigma}(t,t')=i\langle d_{j,\sigma}^\dagger (t')
d_{j,\sigma}(t) \rangle $, $G^{r}_{j\sigma}(t,t')=-i\theta
(t-t')\langle \{ d_{j,\sigma}(t),d_{j,\sigma}^\dagger (t') \}
\rangle $, and $G^{a}_{j\sigma}(t,t')=i\theta (t'-t)\langle \{
d_{j,\sigma}(t),d_{j,\sigma}^\dagger (t') \}  \rangle $,
respectively. These one-particle Green's functions are related
recursively to other Green's functions and correlation functions via
a hierarchy of equations of motion (EOM). To clarify the nonlinear
thermal behavior of EHE in the Coulomb blockade regime, we
numerically solve Eqs. (3) and (4) by considering all correlation
functions and Green's functions.[12,13] To design an EHE driven by
an applied temperature-bias, the thermal induced voltage
($eV_{th}=\mu_L-\mu_R$) across the external load with conductance
$G_{ext}$ needs to be calculated for a given temperature bias
$\Delta T$. To obtain $eV_{th}$, we have to solve self-consistently
all correlation functions appearing in Eq.~(3) subject to the
condition $G_{ext}V_{th}+J=0$, where $J=(J_{L}+J_{R})/2$ is the net
charge current. The heat current satisfies the condition
$Q_L+Q_R=-J* V_{th}$, which denotes the work done by the EHE per
unit time. The efficiency of EHE is thus given by
\begin{equation}
\eta=|J*V_{th}|/Q_L.
\end{equation}

To reveal the importance of many-body effect, which is fully
accounted for in the numerical method,[12,13] we rewrite the net
charge and heat currents as[16]
\begin{eqnarray}
J=\frac{e}{h}\int d\epsilon {\cal T}_{LR}(\epsilon)
[f_L(\epsilon)-f_R(\epsilon)],
\end{eqnarray}
and
\begin{equation}
{\color{red}Q_{L/R}}=\pm \frac{1}{h}\int d\epsilon
~(\epsilon-\mu_{L(R)}){\cal T}_{LR}(\epsilon)
[f_L(\epsilon)-f_R(\epsilon)].
\end{equation}

Notation ${\cal T}_{LR}(\epsilon)$ is the transmission coefficient
for electron transport through the QDMs. Because there are four
possible states for each QD level (empty, one spin-up electron, one
spin-down electron, and two electrons), ${\cal T}_{LR}(\epsilon)$
contains $4^3=64$ configurations for the SCTQD. The analytical
expression of ${\cal T}_{LR}(\epsilon)$ can be found in [16],in
which only one-particle occupation numbers and two-particle on-site
correlation functions used in the Green's functions are considered.
We shall demonstrate that the method of [16] is a good approximation
for QDMs in the low-filling regime.

\section{Results and discussion}

Figure~1(a) shows the total occupation number ($N_t=\sum_{\sigma}
(\lan n_{L\sigma}\ran+\lan n_{C\sigma}\ran+\lan n_{R\sigma}\ran $ )
of SCTQD without thermal bias ($k_B\Delta T=0$) as a function of the
applied gate voltage $V_g$ (which can tune the QD energy level
according to $E_{\ell}=E_F+30\Gamma_0-eV_g$) for three different
temperatures ($k_BT_C=1, 3, 5\Gamma_0$). The electric bias is set at
$e\Delta V=-1\Gamma_0$. The staircase behavior of $N_t$ is due to
the charging effect arising from electron intradot and interdot
Coulomb interactions. The plateaus of $N_t$ correspond to numbers of
integer charges in the SCTQD as electrons fill the QD levels in the
Coulomb blockade regime. The average occupancies in the center dot
($\lan n_{C\sigma}\ran=N_{C,\sigma}$) and outer dots ($\lan
n_{L\sigma}\ran (N_{L,\sigma}) =\lan n_{R\sigma}\ran
(N_{R,\sigma})$) are also plotted in Fig.~1(a) as dash-double-dots
and dash-dotted curves, respectively. Because of symmetry, the
average occupancies in two outer dots remain the same as $V_g$
varies, which leads to a jump of 2 for $N_t$ for the first two
steps. The corresponding tunneling currents $J_{\Delta V}$ are
plotted in Fig.~1(b). The negative sign of $J_{\Delta V}$ indicates
that charge carriers are flowing from the right electrode to the
left electrode. The tunneling currents are appreciable only in the
regions where $N_t$ jumps a step, but become blocked when $N_t$ is
flat as a function of $eV_g$. In the absence of electron Coulomb
interactions, there are three resonant channels of
$\epsilon=E_0-\sqrt{2}t_c$, $\epsilon=E_0$ and
$\epsilon=E_0+\sqrt{2}t_c$. When
$\epsilon_1=E_0-\sqrt{2}t_c=E_F+30\Gamma_0-\sqrt{2}t_c$ is aligned
with the $E_F$ of electrodes, we reach the maximum of the
$\epsilon_1$ peak. Once $E_L=E_R$ are well below $E_F$, the central
QD is depleted (see the curve labeled by $\lan n_{C\sigma}\ran$).
Meanwhile,the outer QDs are filled with one electron for each QD
($\lan n_{L\sigma}\ran+\lan n_{R\sigma}\ran=1$). The situation
remains unchanged until the energy level of $\epsilon_2\approx
E_C+U_{LC}+U_{CR}$ is aligned with $E_F$, one electron is filled
into the central QD. The peak of $\epsilon_2$ describes the
three-electron process. For example, one electron with spin up
(down) of the left electrode tunnels into the left QD with spin down
(up) via the energy level of $E_C+U_{LC}+U_{CR}$ and transfer to the
right QD with spin down (up). Such a three-electron process is
blockaded with increasing the gate voltage. When $E_L+U_L$ and
$E_R+U_R$ are below $E_F$, the increasing two electron occupation
probability weight of outer QDs
($\sum_{\sigma}(N_{L,\sigma}+N_{R,\sigma})=4$) suppresses the
probability of three-electron process. Although one electron is
injected into the central QD when $E_C+2U_{LC}+2U_{CR}$ is aligned
with $E_F$, the transport probability of this five electrons of
SCTQD molecule is extremely small due to ($E_C+2U_{LC}+2U_{CR}$)not
line up with $E_L+U_L+U_{LC}$ and $E_R+U_R+U_{CR}$. Note that these
plateaus are washed out with increasing temperature. The
Hartree-Fock approximation method widely used for molecular
junctions can not reveal the charge transport through molecules in
the Coulomb blockade regime.[10,11] The maximum currents prefers the
orbital-depletion regime of SCTQDs molecule. $J_{max}$ is suppressed
with increasing temperature ($T_C$). Fig. 1(c) shows the charge
current driven by a temperature bias for various  values of $k_BT_C$
with $\Delta V=0$. Unlike $J_{\Delta V}$, $J_{\Delta T}$ shows the
bipolar Coulomb oscillatory behavior with respect to $eV_g$.
Positive (negative) sign indicates that $J_{\Delta T}$ is from the
left (right) electrode to the right (left) electrode. When QD energy
levels are above $E_F$, electrons of the left (hot) electrode
diffuse into the right (cold) electrode by a temperature bias. On
the other hand, electrons of the cold electrode can diffuse into the
hot electrode when QD energy levels are below $E_F$. In general, we
introduce "the hole picture", which is defined as the states below
$E_F$ without electron occupation, to illustrate the behavior of
negative $J_{\Delta T}$. When electron hole balances, $J_{\Delta T}$
vanishes. The sign change of $J_{\Delta T}$ as $V_g$ varies
indicates a bipolar effect. Such a behavior is very different from
the charge current driven by an applied bias $e\Delta V$. It is
worth noting that the maximum $J_{\Delta T}$ is suppressed with
increasing $T_C$. Recently, the bipolar behavior of $J_{\Delta T}$
was experimentally reported in a single metallic QD junction
system.[17]

In the operation of EHE, a temperature bias $\Delta T$ should induce
a thermal voltage $V_{th}$ ($ eV_{th}=\mu_L-\mu_R$) which depends on
the load conductance $G_{ext}$. Fig. 2(a) shows the charge current
($J$) driven by a temperature bias, $k_B\Delta T=1\Gamma_0$ at
various values of $k_BT_C$. We see that the behavior of charge
current shown in {\color{red}Fig. ~2(a)} is similar to that of
Fig.~1(c) with zero load resistance (i.e $1/G_{ext}=0$), which
display a bipolar Coulomb oscillatory behavior. Fig.~2(b) shows the
thermal voltage $V_{th}$ induced by $\Delta T$. This thermal voltage
has two kinds of characteristics. The sign of $eV_{th}$ is opposite
to that of $J$. Meanwhile, the magnitude of $V_{th}$ is in
proportion to $J$. This counter active behavior of $V_{th}$ and $J$
is related the Lenz's law in TE effect. Fig.~2(c) shows the the EHE
efficiency, $\eta$ for various values of $T_C$. It is seen that the
peak values of $\eta$ is suppressed as $T_C/T_H$ increases and
$\eta$ reduces to zero as $T_C/T_H$ approaches 1 (i.e $\Delta T=0$).
Such a behavior is similar to a Carnot engine or an ideal TE device
(with $ZT$ approaching infinity), for which the efficiency is given
by $\eta_C=(1-T_C/T_H)$.[1,2] From the results of Fig.~2, we see
that the highest efficiency of EHE occurs near the transition where
$N_t$ goes from 0 to 1 (with $eV_g\approx 25 \Gamma_0$), which is in
the low-filling regime. When QD energy levels are below $E_F$, not
only the charge current but also the EHE efficiency is suppressed
owing to the strong electron correlation. The EHE of a single QD
with one energy level was theoretically discussed without
considering $V_{th}$ and electron Coulomb interactions in references
[18-19]. The approach considered in references [18-20] is similar to
the case discussed in Fig.~1, where $\Delta T$ and $\Delta V$ are
unrelated. Based on the approach of references of [18-20], the
Lenz's law will not apply.

To reveal the importance of electron correlation arising from many
body effect, the physical quantities of Fig.~2 are recalculated by
Eqs. (6) and (7), where for the transmission factor, ${\cal
T}_{LR}(\epsilon)$ we include only the one-particle occupation
number for each QD and intradot two-particle correlation
functions[16]. The resulting curves are shown in Fig.~3, which have
one-to-one correspondence to those of Fig.~2. For the low-filling
situation (with $eV_g < 30 \Gamma_0$), the results agree very well
with the full-calculation results shown in Fig.~2. On the other
hand, there are appreciable differences between the two results as
$N_t$ exceeds 1 (with $eV_g > 30\Gamma_0$), although their behaviors
are qualitatively the same for $eV_g$ up $100\Gamma_0$. This implies
that a simplified model without considering interdot correlation
functions is sufficient to model the main characteristics of the EHE
made of SCTQDs in the low-filling regime ($N_t\le 1$).

It is difficult to analyze the physical mechanisms for the charge
current given in Eq.~(3) in the nonlinear regime. In stead, we can
analyze Eq. (6) in the linear response regime, where we have
$J=J_{\Delta V_{th}}+J_{\Delta T}={\cal L}_0\Delta V_{th}+{\cal
L}_1\Delta T$. The charge current now has two driving forces, namely
$\Delta V_{th}$ and $\Delta T$. The thermoelectric coefficient
${\cal L}_n$ can be expressed as

\begin{equation}{\cal L}_n=\frac{2e^2}{h}\int d\epsilon {\cal
T}_{LR}(\epsilon)(\frac{(\epsilon-E_F)}{eT})^n\frac{\partial
f(\epsilon)}{\partial E_F}, \end{equation}
where $f(\epsilon)=1/(exp^{(\epsilon-E_F)/k_BT}+1)$ is the equilibrium Fermi
distribution function and $T$ denotes the equilibrium temperature of electrodes. $J_{\Delta V_{th}}$ and
$J_{\Delta T}$ for the first resonant channel in the weak-tunneling limit, $\Gamma/k_BT \ll 1$ can be expressed as
\begin{equation}
J_{\Delta V_{th}}=\frac{2e^2}{h}\frac{\pi \Gamma P_1}{k_BT} \frac{4
t^2_{LC}t^2_{CR}}{(t^2_{LC}+t^2_{CR}+\Gamma^2)^2} \frac{\Delta
V_{th}} {cosh^2\frac{E_0-E_F}{2k_BT}}.
\end{equation}

\begin{equation}
J_{\Delta T}=\frac{2e}{h}\frac{\pi \Gamma P_1}{k_BT^2} \frac{4
t^2_{LC}t^2_{CR}(E_0-E_F)}{(t^2_{LC}+t^2_{CR}+\Gamma^2)^2}
\frac{\Delta T} {cosh^2\frac{E_0-E_F}{2k_BT}},
\end{equation}
where
$P_1=(1-N_{L,\bar\sigma})(1-N_{C,\sigma}-N_{C,\bar\sigma}+c_C)(1-N_{R,\sigma}-N_{R,\bar\sigma}+c_R)$
denotes the probability weight of SCTQDs with an empty state, which
is determined by the single particle occupation number ($N_{\ell}$)
and intradot two particle correlation functions ($c_{\ell}$).[16]
From Eqs.~(9) and (10), we see that the maximum $J_{\Delta V_{th}}$
and $J_{\Delta T}$ occur at $t_{LC}=t_{CR}$. Thus, inhomogenous
electron hopping strength will reduce $J$. When we consider an open
circuit ($G_{ext}=0$), {\color{red}the linear Seebeck coefficient
($S=\Delta V_{th}/\Delta T=-(E_0-E_F)/(eT)$) provides the behavior
of $\Delta V_{th}$, which is irrelevant with $t_{\ell,j}$,
$U_{\ell}$, $U_{\ell,j}$ and $\Gamma$.[16]} The bipolar behavior of
Figs.~1(c) and 2(a) can be explained by Eq.~(10).

So far, we have fixed $G_{ext}=0.2 G_0$, where $G_0=2e^2/h$ is the
quantum conductance. The case of $G_{ext}=0$ was studied in our
previous studies for the design of electronic thermal
rectifiers.[7,8] In the inset of Fig. 3, we plot $\eta=|J\times
V_{th}|/(Q_L+Q_{ph})$ versus $V_g$ for four different values of
$G_{ext}$ at $k_BT_c=1\Gamma_0$, $k_B\Delta T=1\Gamma_0$ and
$t_C=1\Gamma_0$. Note that the calculations of results shown in the
inset include the effect of phonon heat flow given by
$Q_{ph}=\kappa_{ph,0}F_s \Delta T$, where
$\kappa_{ph,0}=\frac{\pi^2k^2_B\bar T}{3h}$ is the universal phonon
thermal conductance arising from acoustic phonon confinement in a
nanowire. $F_s=0.1$  when one considers the phonon scattering from
QDs embedded in a nanowire.[16] $\bar T=(T_C+T_H)/2$. The maximum
efficiency is obtained at $G_{ext}=0.05G_0$. Meanwhile, the maximum
$\eta$ for $G_{ext}=0.2G_0$ (blue line) is around 0.08 including the
effect of $Q_{ph}$, which is much smaller than the value obtained
with $Q_{ph}=0$ as shown by the black solid line of Fig. 3.

To understand the effect of $G_{ext}$ shown in the inset, we can
also compare with the $\eta$ derived by classical approach
considered in references[1,2] and obtain
\begin{equation}
\eta=(1-\frac{T_C}{T_H})\frac{m}{m+(1+m)^2/(ZT_H)+\bar T/T_H},
\end{equation}
where $m=G_{e}/G_{ext}$, and $Z=\frac{S^2 G_e}{\kappa}$. $G_e$,
$S=V_{th}/\Delta T$, and $\kappa$ are the {\color{red}internal}
electrical conductance, Seebeck coefficient and thermal conductance
of the TE device. By taking $d\eta/dm|_{m_o}=0$, we obtain
$m_o=G_e/G^o_{ext}=\sqrt{1+Z\bar T}$, which gives the maximum value
of $\eta$. Because $G^{o}_{ext}=G_e/\sqrt{1+Z\bar T}$, $\eta_{max}$
will occur at vanishingly small $G_{ext}$ if $Z\bar T$ becomes very
large, and the limit of Carnot engine with $\eta_{max}=(1-T_c/T_H)$
is reached). When $Z\bar T=2$, $G^0_{ext}=G_e/\sqrt{3}\approx
0.047G_0$ for $G_e=0.08G_0$. In the Coulomb blockaded regime, $G_e$
is much smaller than $G_0$ for $k_B T/\Gamma $ larger than 1.[See
Eq. (9)] The behavior of results shown in the inset of Fig. 3 can be
explained by Eq.~(11).  Previously, we demonstrated that the $ZT$ of
SCQDM can be larger than 2, including the effect of phonon heat
flow.[16]. This implies that QDMs have promising potential for
realizing high-efficiency EHEs.

To further examine the behavior of the EHE efficiency, we plot in
Fig.~4 $J$, $Q_L$ and $\eta$ as functions of $V_g$ for various
values of $k_B\Delta T$ with $T_C$ fixed at $1\Gamma_0$ . We see
that the peak values of $J$, $Q_L$ and $\eta$ all increases with
$\Delta T$. The results of Fig.~4 indicate that a high efficiency
engine with large electrical outputs needs to maintain a high
temperature bias, which in general only exists in systems with high
thermal resistivity (phonon glass). Serially coupled QDs can enhance
the phonon scattering and thus reduce thermal conductivity.
Therefore, a long chain of QD molecules is more suitable than a
short chain for implementing EHE with high efficiency. From the
results of Figs.~(2) and (4), designers can focus on the EHE
operated at the low-filling regime instead of high-filling
situations. In the low-filling regime, one can usually ignore the
interdot Coulomb interactions, whereas the intradot Coulomb
interactions still play an important role for the electron transport
in the Coulomb blockade regime.[13] Because the effect of $Q_{ph}$
is important as shown in the inset in Fig. 3, we also show the
result including the $Q_{ph}$ effect by triangle marks (with
$k_B\Delta T=3\Gamma_0$), which is to be compared with the dotted
line of Fig. 4(c), Obviously, $\eta_{max}$ is suppressed when
$Q_{ph}$ is included.

We have adopted $t_c=1\Gamma_0$ in Figs~(1)-(4). $t_c$ should depend
on the separation between QDs. To clarify the effect of $t_c$ on the
efficiency of EHE, we plot in Fig.~5 the charge current ($J$),
thermal voltage ($V_{th}$), and efficiency ($\eta$) as functions of
$V_g$ for various values of $t_c$. From the expressions of Eq.~(10),
the charge current is proportional to $t^4_c$ in the weak tunneling
limit, $t_c/\Gamma \ll 1$. When $t_c \ge \Gamma$, the maximum of $J$
no longer increases with increasing $t_c$. The behavior of $J$ with
respect to $t_c$ is consistent with the expression of Eq.~(10),
although it is solely valid in the linear response regime. Next, we
see that $V_{th}$ still follows the Lenz's law with respect to $J$.
The results of Fig.~5(c) show that the maximum $\eta$ occurs at
$t_c=1 \Gamma$. Had we not considered the self-consistent solution
of $V_{th}$, the maximum $\eta$ would have occurred at $t_c
\rightarrow 0$.[20] Obviously, the self-consistent treatment of
$V_{th}$ is essential for getting physically meaningful results.
Recently, the thermal voltage yielded by temperature bias for the
metallic coupled QD was experimentally reported at very low
temperatures [21]. Due to metallic coupled QD, the temperature bias
is still in linear response regime.[21] Meanwhile, we note that $J$
always vanishes at $E_0-E_F=0$. This is well illustrated by
Eq.~(10). Due to Lenz's law, $V_{th}$ also vanishes at $E_0=E_F$
(see Fig.~5(b)).

When there is size/shape variation in serially coupled QDs, the
energy level fluctuation (ELF) of QDs will cause a significant
effect on the ZT values.[16] Therefore, it is desirable to examine
the ELF effect on the charge current and EHE efficiency of SCTQD.
The effects of ELF at different sites of SCTQD are shown in Fig.~6.
It can be seen from Fig.~6 that the charge current reduces quickly
as the difference in QD energy levels becomes larger than the
coupling, $t_c$. We found that the effect of ELF for the central QD
is smaller than that for outer QD. Meanwhile, the temperature effect
shown in Fig.~6(a) is very different from the results shown in
Figs.~1 and 2, where the peak width increases significantly with
increasing temperature $T_C$. The results of Fig.~6(a) imply that
the width of peaks depends on parameters such as tunneling rates and
electron hopping strengths, but not on $T_C$. Such behavior can be
understood by the long distance coherent tunneling effect
(LDCT).[12,13,22] When $E_C \ne E_L=E_R$, it will introduce an
effect hopping strength $t_{eff}=-t_{LC}t_{CR}/(E_C-E_R)$ between
the outer QDs. The behavior of curves shown in Fig.~6(b) is called
the nonthermal broadening effect as reported for the DQD case, where
the peak width depends only on the tunneling rate.[23] Such
characteristics can be used to determine the coupling strength
between QDs and electrodes. They are also useful for applications in
low temperature filters.[23]

Although the EHE efficiency is suppressed by  ELF in SCTQDs, such
phenomenon can be used to design an engine with direction-dependent
electrical output. In Figs.~7(a) and 7(b), the charge current ($J$)
and thermal voltage ($V_{th}$) are calculated for an SCTQD with
{\color{red} the staircase energy levels of $E_L=E_R+2\Delta$,
$E_C=E_R+\Delta$ and $E_R=E_F+10\Gamma_0$, where $\Delta$ is the QD
energy level difference}. In our calculations, the change of outer
QD energy levels arising from $V_{th}$ has been included. Namely,
the outer QD levels become $\epsilon_{L(R)}=E_{L(R)}\pm D_\eta
V_{th}$. It's worth noting that the tunable factor $D_{\eta}=0.3$ is
mainly determined by the QD separation.[22] For $\Delta T
> 0$, $T_H$ is on the left electrode. For $\Delta T < 0$, the two
sides are swapped. (See insets of Fig.~7(a)) The forward (backward)
currents ($J_{F(B)}$) are positive (negative), while $V_{th}$ has
opposite sign with respect to $J$. Thus, the Lenz's law between $J$
and $V_{th}$ is maintained. For both forward and backward currents,
the charge currents have a nonlinear dependence on $\Delta T$. With
increasing $\Delta$, the charge currents (or electrical powers) are
suppressed in the wide temperature bias regime. For $\Delta=0$
(solid black curve), the charge currents shows no directionality,
while for $\Delta=2\Gamma_0$ the direction-dependent charge current
becomes apparent. This directionality of charge current can be
qualitatively explained as follows. When $\Delta T > 0$,
$\epsilon_{L}$ and $\epsilon_R$ become aligned with $E_C$ as
$eV_{th}$ changes to around $-2\Gamma_0$, while for $\Delta T < 0$,
$\epsilon_L$ and $\epsilon_R$ are tuned further away from $E_C$.
Therefore, QD energy level shift due to the thermal voltage induced
by the temperature-bias can play a remarkable role for the current
rectification effect in SCTQD with staircase-like energy levels.

Let's define the charge current rectification efficiency as
$\eta_R=(J_F-|J_B|)|/(J_F+|J_B|)$, which is irrelevant to heat
flows. The calculated $\eta_R$ as a function of temperature bias
under various conditions is shown in Fig.~8. Figure~8(a) shows
$\eta_R$ for various values of {\color{red}$\Delta$} with
$t_C=3\Gamma_0$. We see that the highest rectification occurs when
$\Delta=2\Gamma_0$ with $\eta_R$ approaching 0.2 at the high $\Delta
T$ limit. The rectification efficiency actually becomes poorer if
$\Delta$ is too large. Unlike the case with $\Delta=2\Gamma_0$,
$\eta_R$ decreases with increasing $\Delta T$ for
{\color{red}$\Delta=4$ and $6\Gamma_0$.} To reveal the electron
correlation effects, we also calculate $\eta_R$ with the simplified
procedure as described in Ref.[16] and plot the corresponding curves
with triangle marks in Fig.~8(a). It is found that the rectification
efficiency over estimated in the simplified model. Figure~8(b) shows
$\eta_R$ at $\Delta=2\Gamma_0$ for different electron hopping
strengths ($t_C=0.5, 1,$ and $2 \Gamma_0$). $\eta_R$ is found to be
largest for $t_c=2\Gamma_0$ (dotted line), which is also larger than
that for $t_c=3\Gamma_0$ as shown in Fig.~8(a). Thus, the charge
current rectification efficiency is not a monotonic function of
$t_c$, which is similar to that of Fig. 5(c). In Fig.~8(c), we
consider the effect of varying the temperature of the cold side,
$T_C$. The results indicate that the maximum $\eta_R$ reduces with
increasing $T_C$.

Recently, the nonlinear thermoelectric effects of nanostructures for
developing new applications have been reviewed.[24] For phonon
rectifiers, it is very difficult to realize "phonontronics" due to
large leakage of phonon flow arising from acoustic phonons, which
lack suitable phonon confinement.[4-6,24] The heat rectification
phenomena of electrons can only exist in the very low temperature
regime, because it is seriously suppressed by phonon flows.[7-9,25]
On the other hand, the charge current rectification shown in Fig.~7
will be unaffected by the phonon flow. Therefore an EHE made of
serially coupled QDs with direction-dependent electrical current may
prove useful in the advancement of {nonlinear thermoelectric
devices.[26]}

\section{Summary}
The {charge and heat} transport through SCTQD driven by a
temperature-bias is theoretically studied for the application of EHE
which convert the thermal energies into electrical power. This study
clarifies the efficiency of EHE by considering the self-consistent
solution of charge current with the condition of
$G_{ext}~V_{th}+J=0$. We have demonstrated that EHE prefers the
SCTQDs with energy levels above the Fermi energy of electrodes
(orbital depletion situation). The maximum efficiency of EHE is not
necessary to occur at the maximum electrical output. We found that
$J$ is degraded by the position-dependent QD ELF, which may arise
from QD size fluctuation or energy level shift resulting from
$V_{th}$. $\eta_{max}$ of EHE is seriously suppressed in the
presence of phonon thermal conductance. QDMs have promising
potential for realizing high-efficy EHEs due to their low phonon
thermal conductance. The direction-dependent charge current is
illustrated by the SCTQDs with staircase energy levels. The thermal
voltage yielded by a temperature bias plays a remarkable role to
design a engine with directionality driven by a temperature bias.
This study can be extended to the reversed process of Seebeck effect
(nonlinear behavior of Peltier effect) for the application of
nanoscale coolers.



\begin{flushleft}
{\bf Acknowledgments}\\
\end{flushleft}
This work was supported by the National Science Council of the
Republic of China under Contract Nos. MOST 103-2112-M-008-009-MY3 and MOST 104-2112-M-001-009-MY2.

\mbox{}\\
E-mail address: mtkuo@ee.ncu.edu.tw\\
E-mail address: yiachang@gate.sinica.edu.tw\\

\setcounter{section}{0}

\renewcommand{\theequation}{\mbox{A.\arabic{equation}}} 
\setcounter{equation}{0} 

\mbox{}\\

\newpage
\clearpage

\begin{figure}[h]
\centering
\includegraphics[angle=-90,scale=0.3]{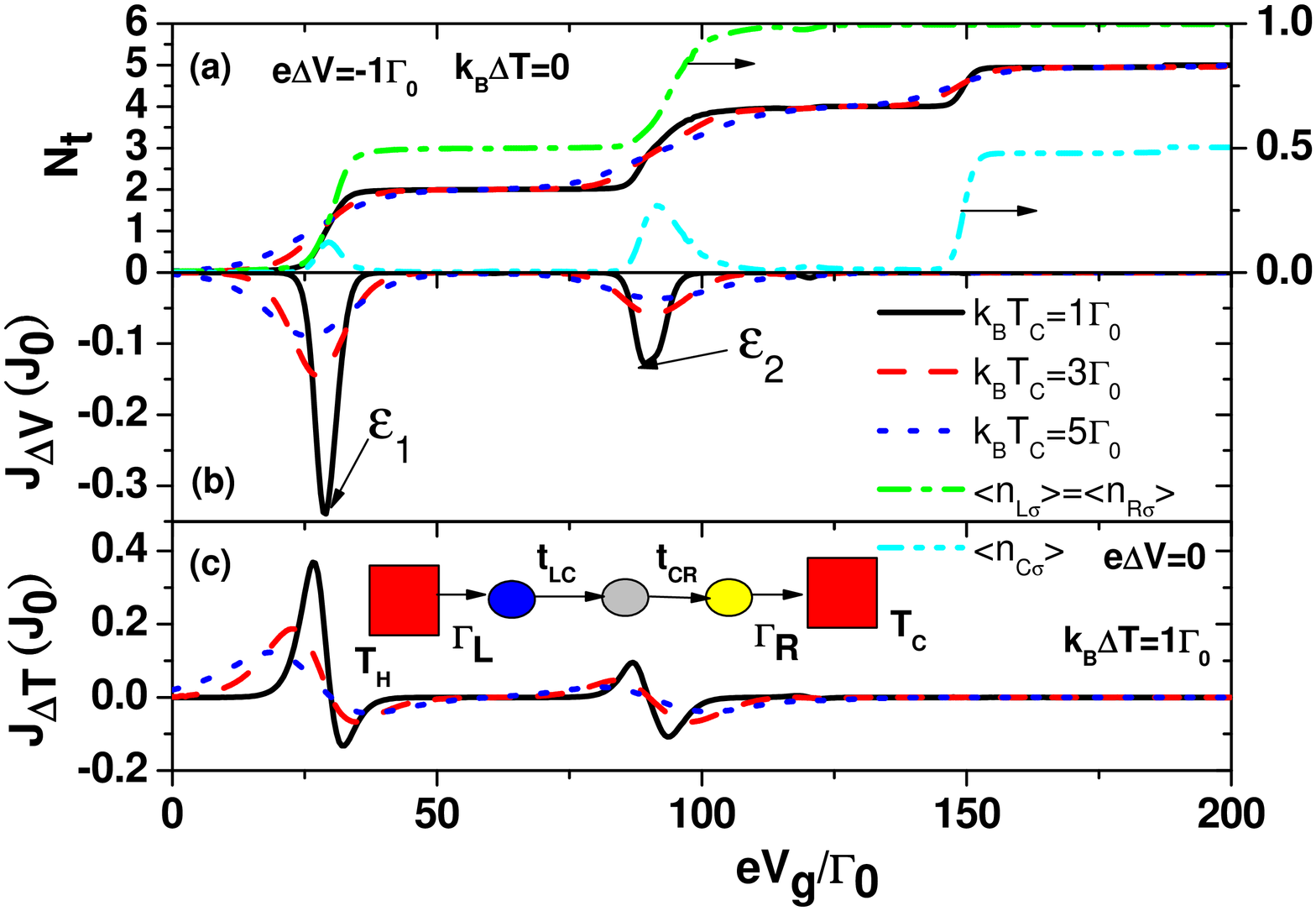}
\caption{(a) Total occupation number, (b) charge current ($J_{\Delta
V}$) at $k_B\Delta V=-1\Gamma_0$ and $k_B\Delta T=0$ , and (c)
charge current ($J_{\Delta T}$) at $e\Delta T=1\Gamma_0$  and
$k_B\Delta V=0$ as a function of quantum dot energy levels
($E_{\ell}=E_0=E_F+30\Gamma_0-eV_g$) for different $T_C$
temperatures. We have following physical parameters
$t_{LC}=t_{CR}=1\Gamma_0$, $t_{LR}=0$, $U_{\ell}=60\Gamma_0$,
$U_{LC}=U_{CR}=30\Gamma_0$, and
$\Gamma_L=\Gamma_R=\Gamma=1\Gamma_0$. $J_{0}=e\Gamma_0/h$}
\end{figure}

\begin{figure}[h]
\centering
\includegraphics[angle=-90,scale=0.3]{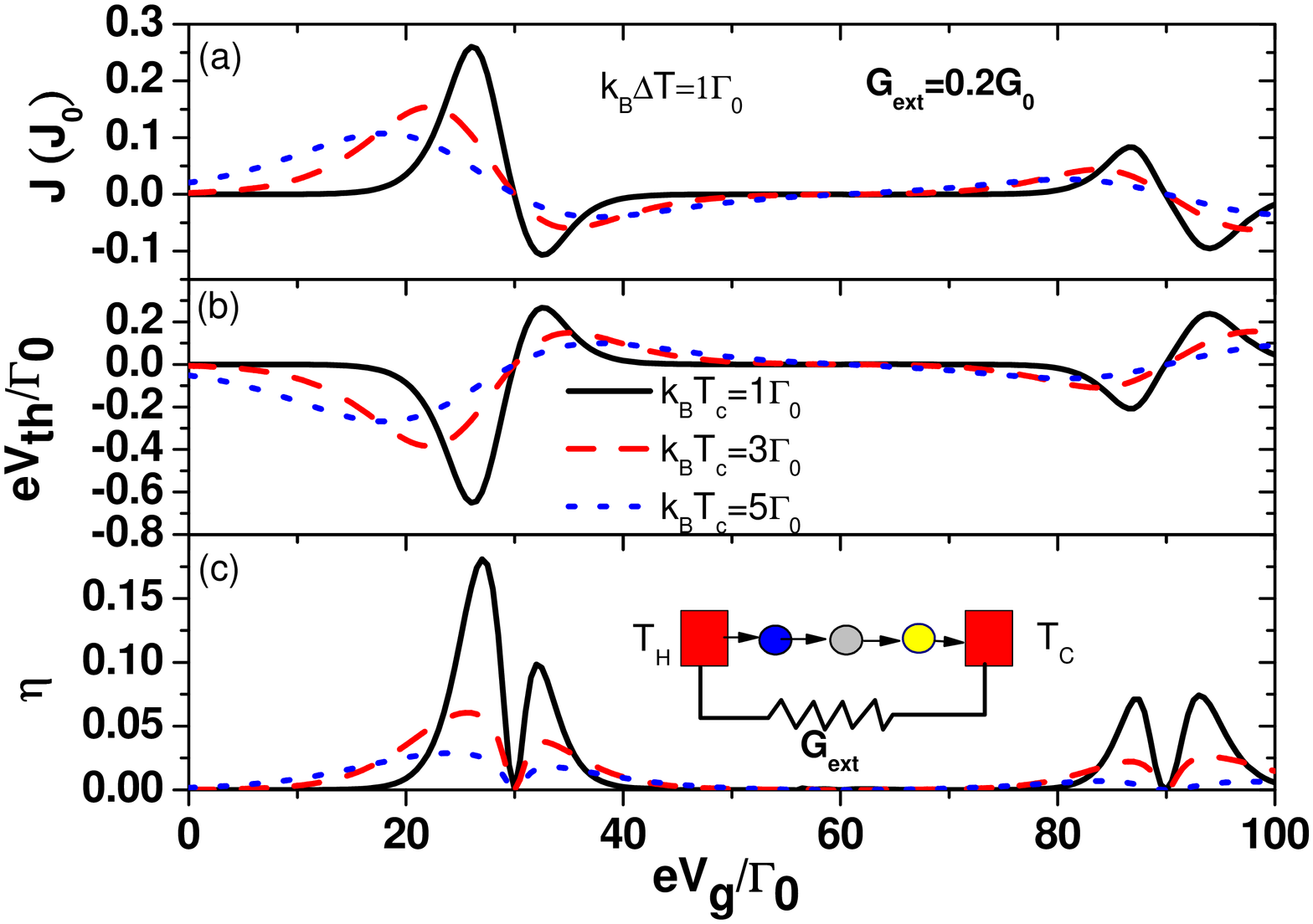}
\caption{(a)Charge current ($J$), (b) thermal voltage ($eV_{th}$)
and (c) efficiency ($\eta$) as a function of QD energy level
($E_{\ell}=E_F+30\Gamma_0-eV_g$) for different $T_C$ values at
$k_B\Delta T=1\Gamma_0$. Other physical parameters are the same as
those of Fig. 1. Other physical parameters are the same as those of
Fig.1. $G_{ext}=0.2 G_0$, where $G_0=2e^2_0/h$.}
\end{figure}

\begin{figure}[h]
\centering
\includegraphics[angle=-90,scale=0.3]{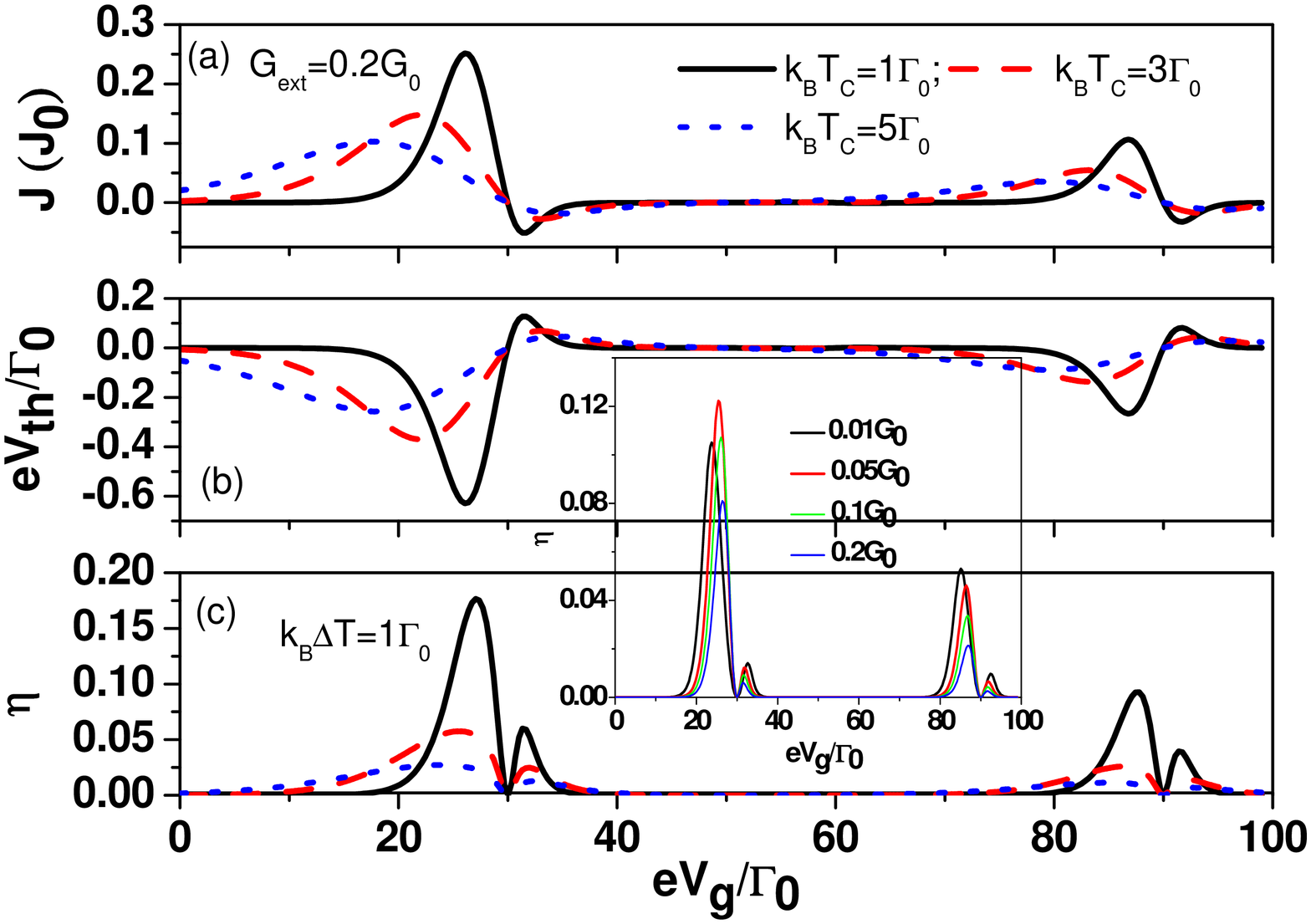}
\caption{(a)Charge current ($J$), (b) thermal voltage ($eV_{th}$)
and (c) efficiency ($\eta$) as a function of QD energy level
($E_{\ell}=E_F+30\Gamma_0-eV_g$) for different $T_C$ values at
$k_B\Delta T=1\Gamma_0$. The curves of Fig. 3 are one to one
corresponding to those of Fig.2. The inset of Fig. 3 shows the
$\eta$ for four $G_{ext}$ values; $0.01,0.05,0.1$ and $0.2G_0$ at
$k_BT_C=1\Gamma_0$.}
\end{figure}

\begin{figure}[h]
\centering
\includegraphics[angle=-90,scale=0.3]{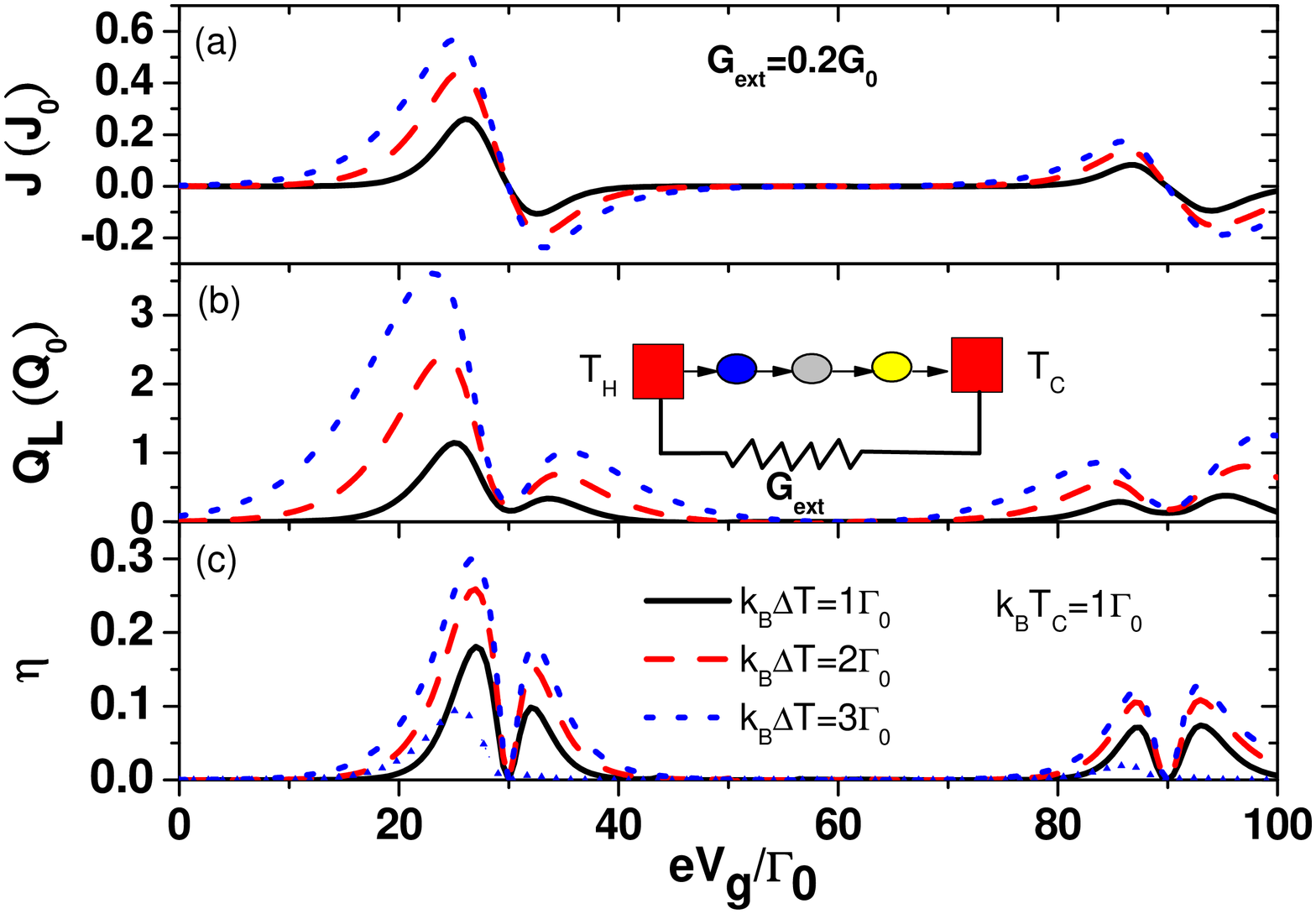}
\caption{(a) Charge current ($J$), (b) heat current ($Q_L$) and (c)
efficiency $\eta$ as a function of QD energy level for different
$k_B\Delta T$ values at $T_C=1\Gamma_0$. We have heat flow in units
of $Q_0=\Gamma_0^2/h$. Other physical parameters are the same as
those of Fig. 1.}
\end{figure}

\begin{figure}[h]
\centering
\includegraphics[angle=-90,scale=0.3]{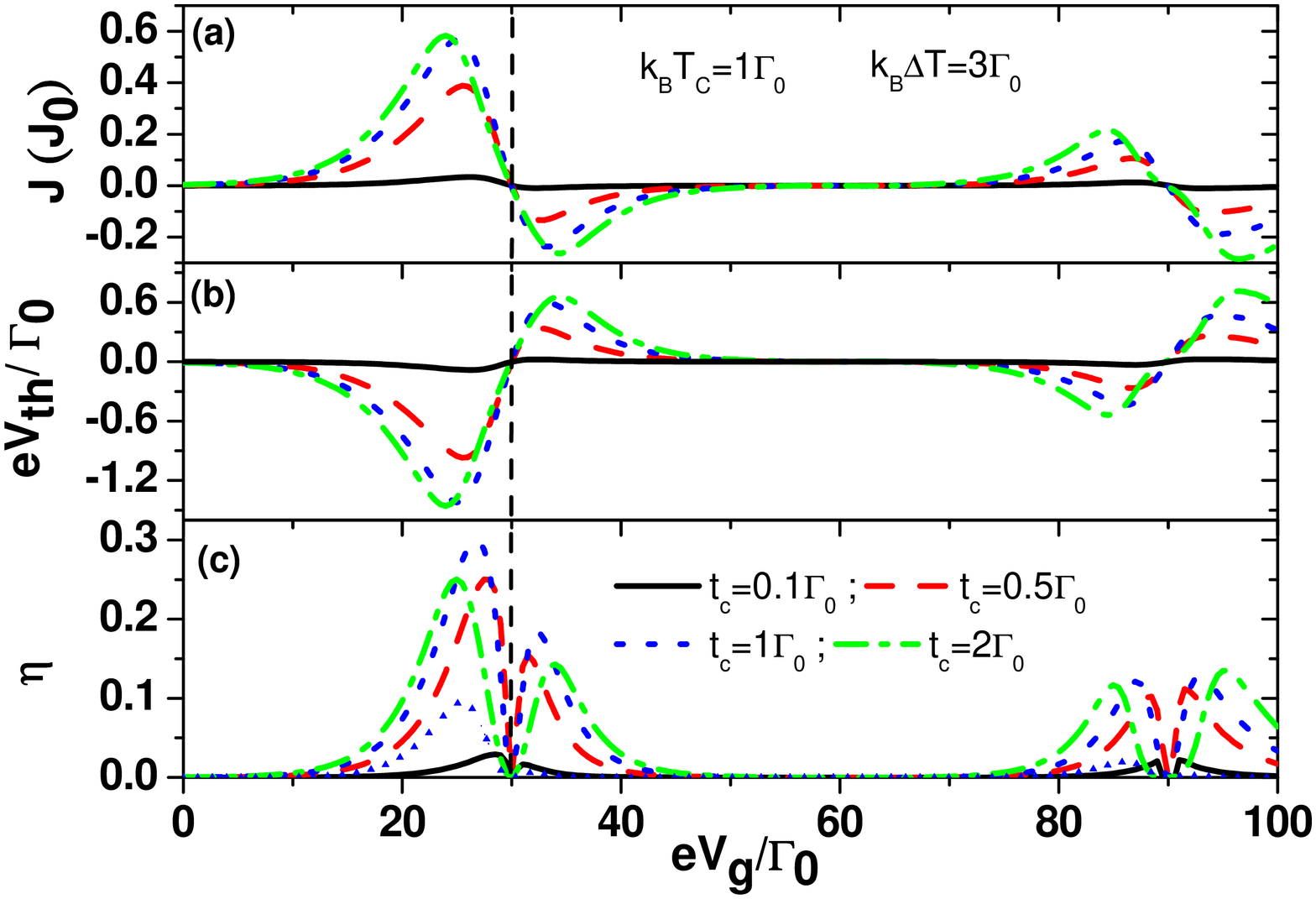}
\caption{(a) Charge current, (b) heat current ($Q_L$) and (c)
efficiency ($\eta$) as a function of QD energy levels for different
$t_c$ values at $k_B\Delta T=3\Gamma_0$, and $k_BT_C=1\Gamma_0$.
Other physical parameters are the same as those of Fig. 1.The curve
with triangle marks is duplicated from Fig. 4(c) to reveal the
$Q_{ph}$ effect.}
\end{figure}

\begin{figure}[h]
\centering
\includegraphics[angle=-90,scale=0.3]{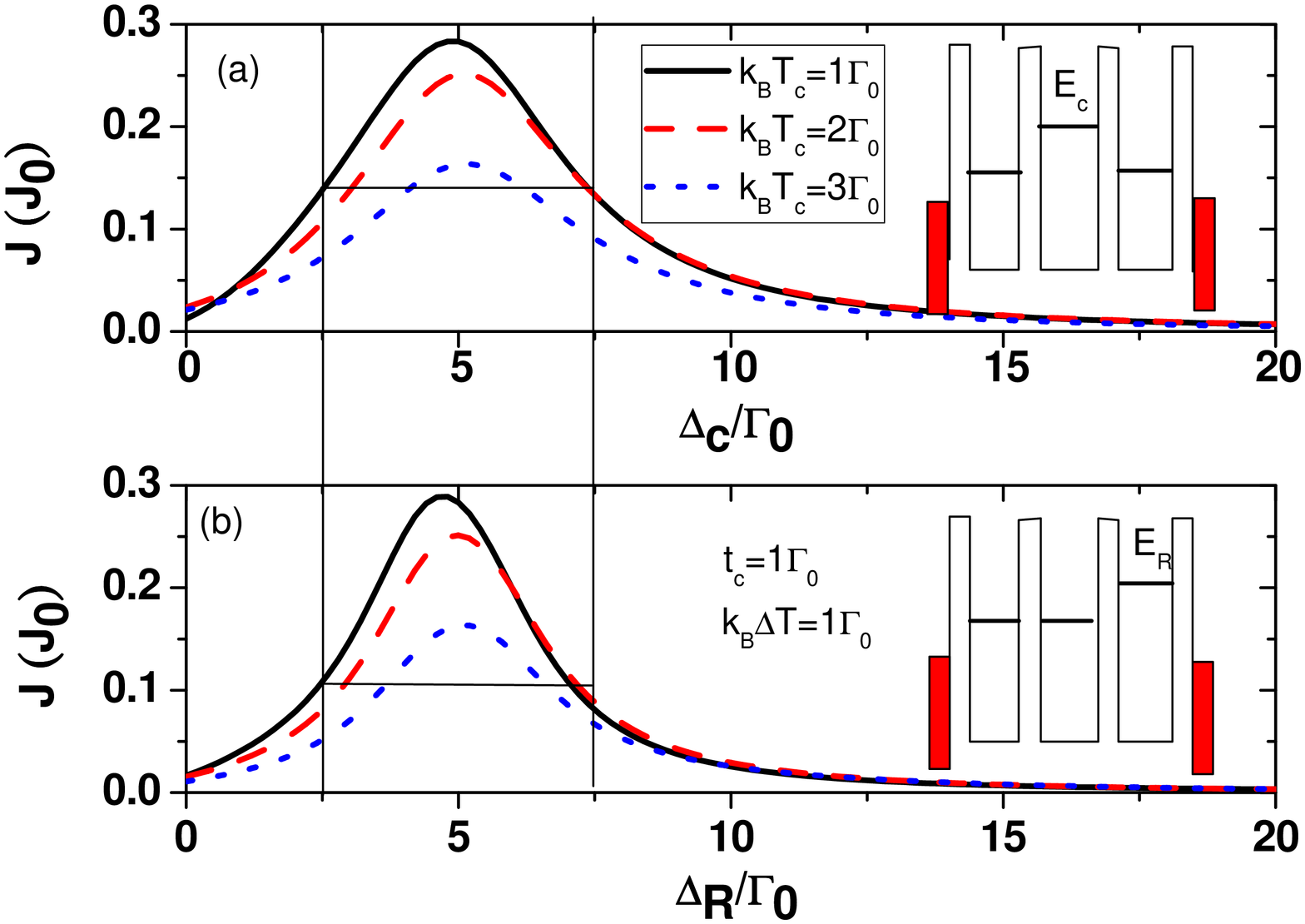}
\caption{Charge current ($J$) as a function of QD energy levels for
different $T_C$ values at $t_c=1\Gamma_0$ and $k_B\Delta
T=1\Gamma_0$. (a) $E_L=E_R=E_F+5\Gamma_0$ and the energy level of center QD ($E_c$) is varied ($\Delta_C=E_C-E_F$).(b)
$E_L=E_C=E_F+5\Gamma_0$ and the energy level of one outer QD ($E_R$) is varied ($\Delta_R=E_R-E_F$). Other physical parameters are the same as those of Fig. 4.}
\end{figure}

\begin{figure}[h]
\centering
\includegraphics[angle=-90,scale=0.3]{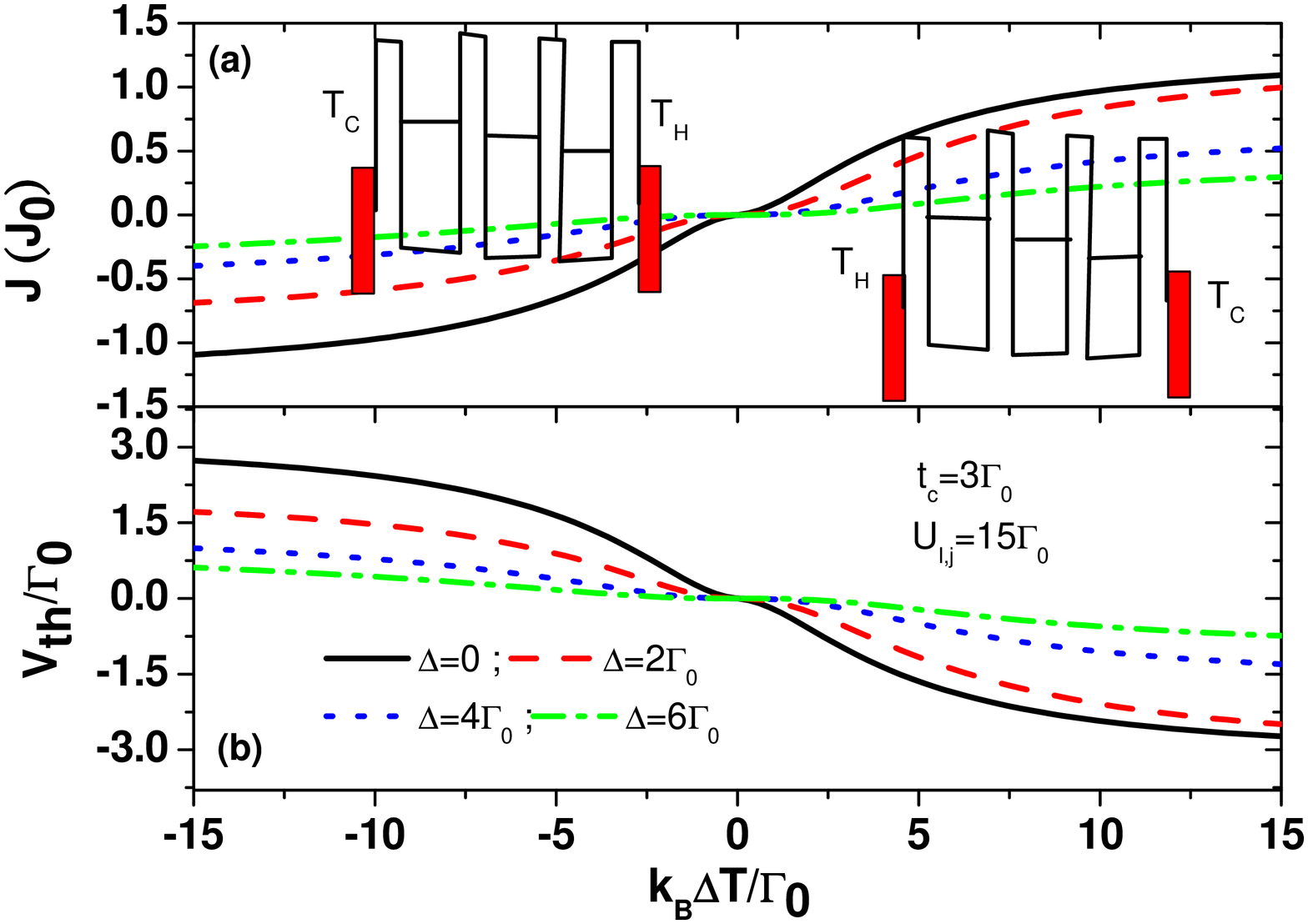}
\caption{(a) Charge current ($J$) and (b) thermal voltage
($V_{th}$) as functions of temperature bias for different QDM
configurations ($E_R=E_F+10\Gamma_0$, $E_C=E_R+\Delta$, and
$E_L=E_R+2\Delta$) with $t_c=3\Gamma_0$, $U_{\ell,j}=15\Gamma_0$, and
$T_c=1\Gamma_0$. We have considered QD energy levels shifted by the
$V_{th}$. Here $E_{L(R)}$ is replaced by
$\epsilon_{L(R)}=E_{L(R)}\pm 0.3V_{th}$. Other physical parameters
are the same as those of Fig. 1.}
\end{figure}

\begin{figure}[h]
\centering
\includegraphics[angle=-90,scale=0.3]{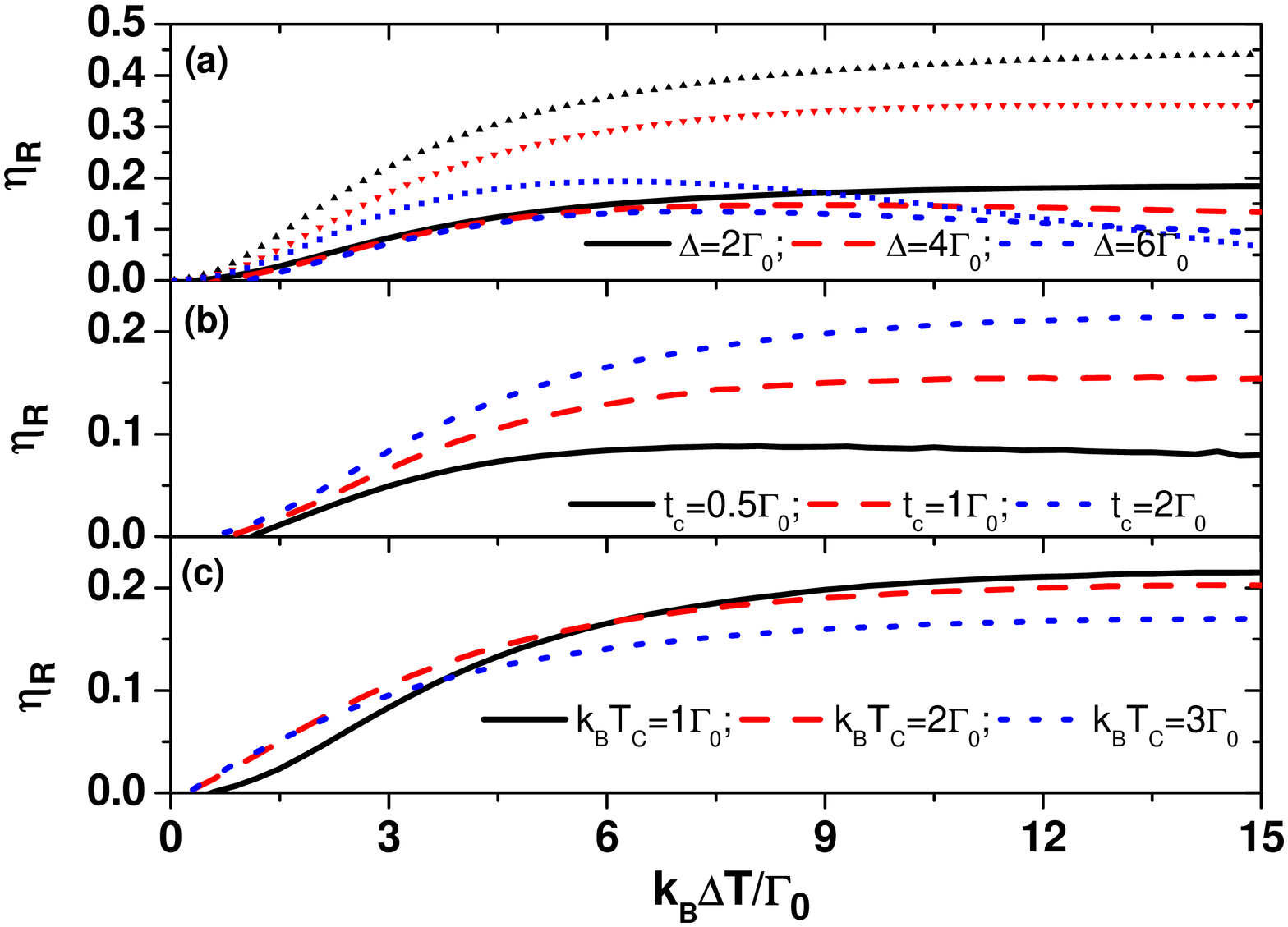}
\caption{ Charge current rectification efficiency ($\eta_R$)  as a
function of temperature bias for the variations of different
physical parameters; (a) $\Delta $ values at $t_c=3\Gamma_0$, and
$T_c=1\Gamma_0$, (b) $t_c$ values at $\Delta=2\Gamma_0$ and
$T_c=1\Gamma_0$, and (c) $T_c$ values at $\Delta=2\Gamma_0$ and
$t_c=2\Gamma_0$. Other physical parameters are the same as those of
Fig. 7.}
\end{figure}

\clearpage


\begin{thebibliography}{50}

\bibitem[1]{Minn} A. J. Minnich,M. S. Dresselhau, Z. F. Ren, and G. Chen,
Energy Environ, Sci. \textbf{2}, 466 (2009).

\bibitem[2]{Zeb} M. Zebarjadi, K. Esfarjania, M.S. Dresselhaus, Z.F. Ren and G.
Chen, Energy Environ Sci \textbf{5},  5147 (2012).

\bibitem[3]{Harm} T. C. Harman, P. J. Taylor,  M. P. Walsh, B. E.
LaForge, Science  \textbf{297}, 2229 (2002).

\bibitem[4]{Chan} C. W. Chang, D. Okawa, A. Majumdar, and Zettl A,
Science\textbf{ 314}, 1121 (2006).

\bibitem[5]{Li} B. W. Li, L. Wang and G. Casati, Phys. Rev. Lett,
Phys. Rev. Lett. \textbf{93}, 184301 (2004).

\bibitem[6]{Hu} B. Hu, L. Yang, and Y. Zhang, Phys. Rev. Lett.\textbf{ 97},
124302 (2006).

\bibitem[7]{Kuo1} D. M. T. Kuo and  Y. C. Chang,
Phys. Rev. B \textbf{81}, 205321 (2010).

\bibitem[8]{Tseng} Y. C. Tseng, D. M. T. Kuo, Y. C. Chang and Y. T. Lin,
Appl. Phys. Letts. \textbf{103},  053108 (2013).

\bibitem[9]{Pere} MJ. Martinez, A. Fornieri and F. Giazotto,
Nature Nanotechnology, \textbf{10}, 303 (2015).


\bibitem[10]{Ber1} J. P. Bergfield and C. A. Stafford, Nano Letters
\textbf{9}, 3072 (2009).

\bibitem[11]{Ber2} J. P. Bergfield, M. A. Solis, and C. A.
Staffford, ACS Nano \textbf{4}, 5314 (2010).


\bibitem[12]{Che1} C. C. Chen, Y. C. Chang and David M T Kuo,
Phys. Chem. Chem. Phys. \textbf{17}, 6606 (2015).

\bibitem[13]{Che2} C. C. Chen, David M T Kuo and Y. C. Chang,
Phys. Chem. Chem. Phys. \textbf{17}, 19386 (2015).









\bibitem[14]{Hau} H. Haug and A. P. Jauho, Quantum Kinetics in Transport and Optics
of Semiconductors (Springer, Heidelberg, 1996).

\bibitem[15]{Jau} A. P. Jauho, N. S. Wingreen and Y. Meir, Phys.
Rev.  B \textbf{50}, 5528 (1994), and references therein.

\bibitem[16]{Kuo2} David M. T. Kuo and Y. C. Chang, Nanotechnology, \textbf{24}, 175403 (2013).

\bibitem[17]{Sven} S. F. Svensson, E. A. Hoffmann, N. Nakpathomkun,
P. M. Mu, H. Q. Xu, H. A. Nilsson, D. Sanchez, V. kashcheyevs and H.
Linke, New. J. Phys. \textbf{15}, 105011 (2013).

\bibitem[18]{Nakp} N. Nakpathomkun, H. Q. Xu and H. Linke, Phys.
Rev. B \textbf{82}, 235428 (2010).

\bibitem[19]{Leij} M. Leijnse, M. R. Wegewijs and K. Flensberg,
Phys. Rev. B\textbf{ 82}, 045412 (2010).

\bibitem[20]{Liu} Y. S. Liu, X. F. Yang, X. K. Hong, M. S. Si, F. Chi and Y. Guo. App. Phys. Lett, 103, 093901 (2013).

\bibitem[21]{Thier} H. Thierschmann, F. Arnold, M.
Mittermuller, L. Maier, C. Heyn, W. Hansen, H. Buhmann and L. W
Molenkamp, New. J. Phys. 17, 113003 (2015).

\bibitem[22]{Kuo5}  David M. T. Kuo and Y. C. Chang, Phys Rev. B, \textbf{89}, 115416 (2014).

\bibitem[23]{Kuo4} D. M. T. Kuo, S. Y. Shiau and Y. C. Chang, Phys. Rev. B.\textbf{ 84}, 245303 (2011).


\bibitem[24]{Zhu} J. Zhu, K. Hippalgaonkar, S. Shen, KV Wang, Y.
Abate, S. Lee, J. Wu, X. Yin, A. Majumdar and X. Zhang, Nano Lett,
14, 4867 (2014).

\bibitem[25]{Their} H. Thierschmann, R. Sanchez, B. Sothmann, F. Arnold, C. Heyn, W. Hansen,
, H. Buhmann and L. W. Molenkamp, Nature Nanotechnolgy, 10,
854(2015).

\bibitem[26]{Soth} B. Sothmann, R. Sanchez and A. N. Jordan,
Nanotechnolgy, 26, 032001(2015).





























































\end{thebibliography}
\end{document}